\documentclass[12pt, letterpaper]{article} 

\usepackage[english]{babel} 

\usepackage{microtype} 

\usepackage{amsmath,amsfonts,amsthm} 

\usepackage[svgnames]{xcolor} 

\usepackage[hang, small, labelfont=bf, up, textfont=it]{caption} 

\usepackage{booktabs} 

\usepackage{lastpage} 

\usepackage{graphicx} 

\usepackage{enumitem} 
\setlist{noitemsep} 

\usepackage{sectsty} 
\allsectionsfont{\usefont{OT1}{phv}{b}{n}} 

\usepackage{aas_macros}  

\usepackage{geometry} 

\usepackage[T1]{fontenc} 
\usepackage[utf8]{inputenc} 

\usepackage{XCharter} 

\usepackage{fancyhdr} 
\fancypagestyle{firstpage}{ 
	\fancyhf{}
}

\newcommand{\authorstyle}[1]{{\large\usefont{OT1}{phv}{b}{n}\color{DarkRed}#1}} 

\usepackage{titling} 

\newcommand{\HorRule}{\color{DarkGoldenrod}\rule{\linewidth}{1pt}} 

\pretitle{
	\vspace{-30pt} 
	\HorRule\vspace{10pt} 
	\fontsize{32}{36}\usefont{OT1}{phv}{b}{n}\selectfont 
	\color{DarkRed} 
}

\posttitle{\par\vskip 15pt} 

\preauthor{} 

\postauthor{ 
	\vspace{10pt} 
	\par\HorRule 
	\vspace{20pt} 
}

\usepackage{lettrine} 
\usepackage{fix-cm}	

\newcommand{\initial}[1]{ 
	\lettrine[lines=3,findent=4pt,nindent=0pt]{
		\color{DarkGoldenrod}
		{#1}
	}{}%
}

\usepackage{xstring} 

\newcommand{\lettrineabstract}[1]{
	\StrLeft{#1}{1}[\firstletter] 
	\initial{\firstletter}\textbf{\StrGobbleLeft{#1}{1}} 
}

\usepackage[backend=bibtex,style=science,maxnames=2,natbib=true]{biblatex} 
\AtEveryBibitem{\clearfield{month}\clearfield{archivePrefix}\clearfield{eprint}\clearfield{primaryClass}}
\AtEveryCitekey{\clearfield{month}\clearfield{archivePrefix}\clearfield{eprint}\clearfield{primaryClass}}

\defbibenvironment{bibliography}
  {}
  {}
  {\addspace
   \printtext[labelnumberwidth]{%
     \printfield{prefixnumber}%
     \printfield{labelnumber}}%
   \addhighpenspace}

\usepackage[autostyle=true]{csquotes} 

\title{Keys of a Mission to Uranus or Neptune, the Closest Ice Giants} 

\author{
	\authorstyle{Tristan Guillot$^1$}, 
	\authorstyle{Jonathan Fortney$^2$}, 
	\authorstyle{Emily Rauscher$^3$}, 
	\authorstyle{Mark Marley$^4$} 
	\authorstyle{Vivien Parmentier$^5$}, 
	\authorstyle{Mike Line$^6$}, 
	\authorstyle{Hannah Wakeford$^7$}, 
	\authorstyle{Yohai Kaspi$^8$}, 
	\authorstyle{Ravit Helled$^9$}, 
	\authorstyle{Masahiro Ikoma$^{10}$}, 
	\authorstyle{Heather Knutson$^{11}$}, 
	\authorstyle{Kristen Menou$^{12}$}, 
	\authorstyle{Diana Valencia$^{12}$}, 
	\authorstyle{Daniele Durante$^{13}$}, 
	\authorstyle{Shigeru Ida$^{14}$}, 
	\authorstyle{Scott Bolton$^{15}$}, 
	\authorstyle{Cheng Li$^{16}$}, 
	\authorstyle{Kevin Stevenson$^{17}$}, 
	\authorstyle{Jacob Bean$^{18}$}, 
	\authorstyle{Nicolas Cowan$^{19}$},
	\authorstyle{Mark Hofstadter$^{20}$},
	\authorstyle{Ricardo Hueso$^{21}$},
	\authorstyle{Jeremy Leconte$^{22}$},
	\authorstyle{Liming Li$^{23}$},
	\authorstyle{Christoph Mordasini$^{24}$}, 
	\authorstyle{Olivier Mousis$^{25}$},
	\authorstyle{Nadine Nettelmann$^{26}$},
	\authorstyle{Krista Soderlund$^{27}$},
	\authorstyle{Michael H. Wong$^{16,28}$}
	\newline\newline 
	$^1$Universit\'e C\^ote d'Azur, France, 
	$^2$University of California, Santa Cruz, USA,
	$^3$University of Michigan, USA,
	$^4$NASA Ames Research Center, USA,
	$^5$University of Oxford, UK,
	$^6$Arizona State University, USA,
	$^7$University of Bristol, UK,
	$^8$Weizmann Institute, Israel,
	$^9$University of Zurich, Switzerland,
	$^{10}$University of Tokyo, Japan,
	$^{11}$California Institute of Technology, USA,
	$^{12}$University of Toronto, Canada,
	$^{13}$Universit\'a di Roma, Italy,
	$^{14}$Earth Life Sciences Institute, Japan,
	$^{15}$Southwest Research Institute, USA,
	$^{16}$University of California, Berkeley, USA,
	$^{17}$Johns Hopkins APL, Laurel, MD, USA,
	$^{18}$University of Chicago, USA,
	$^{19}$McGill University, Canada,
	$^{20}$Jet Propulsion Laboratory, USA,
	$^{21}$Universidad del País Vasco (UPV/EHU), Spain,
	$^{22}$Universit\'e de Bordeaux, France,
	$^{23}$Cornell University, USA,
	$^{24}$University of Bern, Switzerland,
	$^{25}$Universit\'e de Marseille, France,
	$^{26}$DLR Berlin, Institut für Planetenforschung, Germany,
	$^{27}$University of Texas at Austin, USA,
	$^{28}$SETI Institute, USA
}

\date{A White Paper for the Decadal Survey of Planetary Sciences and Astrobiology.  --- \today} 

\begin{document}

\maketitle 

\thispagestyle{firstpage} 

\newpage
\setcounter{page}{1}

\lettrineabstract{Uranus and Neptune are the archetypes of "ice giants", a class of planets that may be among the most common in the Galaxy. They hold the keys to understand the atmospheric dynamics and structure of planets with hydrogen atmospheres inside and outside the solar system; however, they are also the last unexplored planets of the Solar System.  Their atmospheres are active and storms are believed to be fueled by methane condensation which is both extremely abundant and occurs at low optical depth. This means that mapping temperature and methane abundance as a function of position and depth will inform us on how convection organizes in an atmosphere with no surface and condensates that are heavier than the surrounding air, a general feature of  giant planets. Owing to the spatial and temporal variability of these atmospheres, an orbiter is required. A probe would provide a reference atmospheric profile to lift ambiguities inherent to remote observations. It would also measure the abundances of noble gases which can be used to reconstruct the history of planet formation in the Solar System. Finally, mapping the planets' gravity and magnetic fields will be essential to constrain their global composition, atmospheric dynamics, structure and evolution. An exploration of Uranus or Neptune will be essential to understand these planets and will also be key to constrain and analyze data obtained at Jupiter, Saturn, and for numerous exoplanets with hydrogen atmospheres.
}


\vspace{-12pt}
\section{Introduction}
\vspace{-12pt}
Admittedly, ``ice giants'' form a yet not well-defined class of planets between a few times the mass of the Earth and a fraction of that of Saturn. Their name comes from the idea that their mass mostly originates from condensed water ice that accreted in protoplanetary disks. The large amount of water ice led them to become more massive than traditional terrestrial planets but yet without accreting so much hydrogen and helium to fall into the realm of the larger "gas giants". This idea is plausible, because water is the most abundant condensable species and certainly the most crucial building block of planet formation\citep{Weidenschilling1977}. However, it is unproven and we do not know whether our own ice giants, Uranus and Neptune, are mostly made of H$_2$O or whether they may be formed of more ``rocks'' (more refractory species) than ``ices'' \citep[e.g.][]{Kunitomo+2018, Helled+2020}. Recently, detailed analysis of the Kepler survey have shown that planets in the ice giant mass regime may be the most abundant class of planets in our Galaxy \citep{Fulton+2017}.

Yet, the ice giants closest to us, Uranus and Neptune, have never been studied by orbiting spacecrafts. 
Contrary to all other planets in the solar system, they have only been visited for a couple of days each and from a distance by the Voyager 2 spacecraft flybys.

\begin{figure*}
	\includegraphics[width=\linewidth]{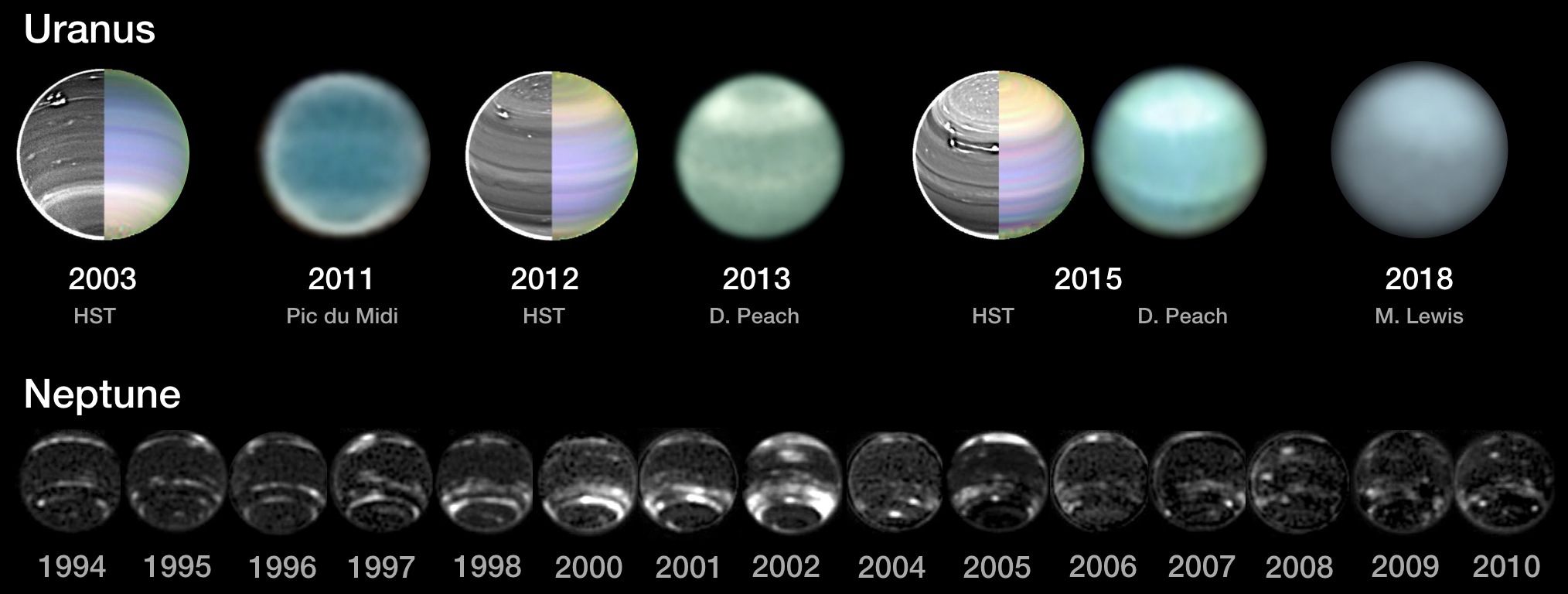} 
\caption{Images of Uranus and Neptune showing seasons and storms. The HST/STIS images of Uranus correspond to H band (left) and false color (right) images \citep{Sromovsky+2019}. Amateur images from the Pic du Midi, D. Peach and M. Lewis have been taken from the PVOL database (http://pvol2.ehu.eus/). The images of Neptune have been obtained from HST/WFPC2 in the visible \citep{Karkoschka2011b}.}
	\label{fig:globes} 
\end{figure*}

Both Uranus and Neptune are fascinating planets that hold some of the keys to understand the origin of our Solar System and to make sense of the observations of exoplanetary atmospheres. As seen in Fig.~\ref{fig:globes}, they both have active, complex atmospheres, observed and monitored by professional and amateurs alike.  We advocate that the exploration of our Solar System must continue and that either Uranus or Neptune, or both, should be the next targets in this journey that will ultimately help us to understand exoplanets as well. 

\vspace{-12pt}
\section{Keys to understanding hydrogen atmospheres}
\vspace{-12pt}
Two major particularities of the atmospheres of giant planets are the absence of a surface and the fact that condensates are heavier than surrounding gas, creating a meteorological regime that is intrinsically different from that of terrestrial planets: Moisture tends to sink instead of rising, and with no surface, it is not clear how deep condensing species will sink. In spite of this, since these planets are convective and storms are regularly observed, the prevailing view has been that this is a minor effect that can be largely ignored: Convective motions should homogenize composition below the condensation level (the ``cloud base'') and latent heat effects should lead to powerful storms capable of an efficient upward transport of condensable species. The Galileo probe measurements \citep{Wong+2004} and the Juno measurements \citep{Bolton+2017, Li+2017} have shown that this view is at best incomplete and perhaps altogether wrong. 

There is now ample evidence that the two major condensing species in Jupiter's atmosphere, water and ammonia, have spatially variable abundances much below their condensation level. In Jupiter, water was found to be subsolar in a hot spot by in situ measurements of the Galileo probe down to at least 20 bar \citep{Wong+2004}, but is also significantly sub-saturated at other locations \citep[e.g.,][]{Roos-Serote+2004}, while it is nearly saturated and at least solar in the Great Red Spot \citep{Bjoraker+2018} and in the equatorial zone \citep{Li+2020}. Ammonia has long been found depleted in large regions of Jupiter down to several bars at least \citep{dePater+1991, dePater+2016}, but it has now been found to be variable much deeper, down to 30 bars or more \citep{Bolton+2017, Li+2017}.  In Saturn, there is also evidence of large-scale latitudinal variations in the ammonia abundance, similar to Jupiter \citep{Fletcher+2011}, potentially influenced by the decade-scale compositional and thermal changes within the intermittent convective cycle \citep{Li+Ingersoll2015}.


Deep variability of volatiles on Jupiter and Saturn is an issue not only for constraining bulk composition, but also for interior and evolutionary models of the entire class of
planets with hydrogen atmospheres. The assumption of a uniform upper boundary for one-dimensional models is largely validated by observed one-bar temperature fluctuations in Jupiter and Saturn of a few percent at most \citep[e.g.][]{Ingersoll+Porco1978, Fletcher+2020}. But what happens deeper is not clear. The abundance variations in ammonia and water indicate that large regions must be on average stable to convection. Storms, in particular water storms, appear to be essential for transporting the interior heat flux \citep{Gierasch+2000}. For large abundances of condensing species, the temperature profile is unknown \citep{Guillot1995, Leconte+2017, Friedson+Gonzales2017}.   Variability in heat transport and cooling raise the possibility that 2D/3D models of the deeper interior may be needed to infer the planet's structure and evolution. 

These issues are not confined to ammonia and water. They extend to any condensing species with an abundance that is large enough to affect energy transfer. Importantly, this is the case of helium which is known to separate from hydrogen at Mbar regions in Jupiter and Saturn \citep[e.g.][]{Stevenson+Salpeter1977, Morales+2013}. 

Understanding how hydrogen atmospheres transport heat and elements is a formidable task. It involves multiple scales, from the global scale (i.e., the size of the planet itself, $\sim 100,000$\,km) to the sizes of storms ($\sim 1-100$ km) and includes complex hydrodynamics and microphysics. Global circulation models (GCMs) \citep[e.g.][]{Dowling+1998, Kaspi+2009, Liu+Schneider2010, Guerlet+2014} are challenging computational endeavors, particularly for planets with deep atmospheres, and thus must simplify the treatment of convective storms and clouds. Cloud or cloud-ensemble models \citep[e.g.][]{Hueso+SanchezLavega2001, Sugiyama+2014, Li+Chen2019} do not include meridional motions and/or global scale winds. Detailed microphysical treatments \citep[e.g.][]{Yair+1995} are based on the Earth's schemes and must be extrapolated to be applied to the giant planets. Therefore, numerical simulations can only guide us on what may be occurring in these atmospheres. We need  ground truth. 

Unfortunately, the measurements required to give new insight are scarce because in Jupiter and Saturn most of the action occurs hidden from view at large optical depth. The ammonia condensation region near 0.7 bar in Jupiter and 1.5 bar in Saturn is observable, but ammonia has a low abundance ($\sim 100$ to $500$\,ppmv mixing ratio) and can only drive a weak moist convection \citep[e.g.][]{Stoker1986}. Instead, most of the storms that we see must be powered by water condensation \citep[see][]{Hueso+SanchezLavega2001, Hueso+2002, Sugiyama+2014, Li+Chen2019}, at levels of $\sim 6$\,bar in Jupiter and $\sim 12$\,bar in Saturn. Juno's MWR instrument was able to probe these regions and deeper in Jupiter but the measurements are mostly sensitive to ammonia's absorption, now believed to be a complex function of depth, latitude and possibly even longitude \citep{Li+2017}. The effect of water is indirect. Finally, we lack a well-defined temperature pressure profile that would allow lifting some of the degeneracies in the measurements. 

Uranus and Neptune possess one key ingredient to understand atmospheric dynamics in hydrogen atmospheres: They are cold enough for methane to condense at low pressure levels $\sim 1.5$\,bar \citep{Lindal1992}, in a region of the troposphere at modest optical depth, and methane is present in abundance to drive moist convection at these levels \citep{Stoker+Toon1989}. Methane is extremely abundant and its abundance is variable with latitude. The maximum mixing ratio in Uranus inferred from HST, Keck and IRTF observations is $f_{\rm CH_4}=2.55\%$ to $3.98\%$ \citep{Sromovsky+2019}. In Neptune, the maximum value detected with VLT/MUSE at a latitude $30^\circ$S is even higher, $f_{\rm CH_4}=5.90\pm 1.07\%$ \citep{Irwin+2019a}. Thus, for both planets, methane accounts for 15\% to 30\% of the mass in the upper atmosphere, higher but comparable to the expected 2\% to 10\% for water in Jupiter and Saturn. The study of methane condensation in Uranus and Neptune can therefore be used to understand moist convection in general, and particularly in this difficult regime where it is inhibited by the molecular weight \citep{Guillot1995}.  

\vspace{-12pt}
\section{Exoplanets: an expanding dataset}
\vspace{-12pt}
We are now about 20-25 years into work characterizing the physics of giant exoplanets.  Transiting giant planets in particular have allowed for an assessment of giant planet thermal evolution, atmospheric composition, and atmospheric dynamics under strong stellar forcing.  The next decade of this science will be truly transformational, with the continuation of \emph{TESS} and the rise of \emph{JWST}, \emph{ARIEL}, \emph{PLATO}, and high-resolution spectrographs on planned Extremely Large Telescopes (ELTs).  Planets between the sizes of Earth and Neptune are the most abundant yet found, and will see a dramatic improvement in their atmospheric characterization.  Models suggest that planets in $\sim2-4$ Earth radius regime harbor a hydrogen-dominated atmosphere \citep{Lopez2014}, and such planets will be excellent targets for atmospheric study.

Observers will work to understand atmospheric circulation, including wind speeds, hot spot offsets, and day-night temperature contrasts.  Atmospheric composition, via spectroscopy of H$_2$O, CO, CO$_2$, CH$_4$, NH$_3$, and a wide variety of atomic metals in the hottest planets, will dramatically alter our view of giant planet atmospheric abundances.  A number of theoretical works have aimed to tie, for instance, C, O, (and N) abundances to the distance of formation within the disk and the relative accretion of solids and gas \citep{Oberg+2011,Piso+2016}. Just like the field of exoplanet structure and bulk composition has moved into the realm of statistical studies of larger samples \citep{Thorngren+2016, Thorngren+Fortney2018}, which will further expand, the same will be true of many aspects of exoplanet atmospheres \citep[e.g.,][]{Zellem+2019}.

However, detailed understanding of the solar system's giant planets provides the only context for these statistical studies.  For instance, there is now clearly tension between the simple internal structure models applied to hot Jupiters \citep[e.g.,][]{Thorngren+2016} and the newest insights from Jupiter/Juno \citep{Wahl+2017} and Saturn/Cassini \citep{Mankovich2020} that suggest dilute cores.  Future exoplanet work will incorporate these lessons, and we see a similar path for detailed knowledge of the atmosphere and interior structure of Uranus and Neptune.


\emph{JWST} and \emph{ARIEL} will lead to far more robust assessments of atmospheric dynamics and abundances, compared to previous efforts with \emph{Spitzer} and \emph{Hubble}.  It seems assured that results from exoplanet phase curves and spectroscopy will show shortcomings in the relatively simple atmospheric models that have been applied to these planets so far.  While the past two decades have seen advances in the characterization of exoplanets close to their star, present and future missions will enable characterizing the atmospheres of cooler planets. Already, this is exemplified by initial observational studies of K2-18 b, a planet with a hydrogen atmosphere possibly containing water vapor in the habitable zone of its star \citep{Tsiaras+2019, Benneke+2019}.  System age will also be a new axis to study.

Direct spectral imaging also allows characterization of exoplanetary atmospheres at greater orbital distances, more similar to the heliocentric distances of our ice giants. With current technology, directly imaged expolanets are young (hot), so characterizing them requires accurate thermal evolution models that benefit from solar system constraints \citep{Marleau+Cumming2014}.

GCMs have been shown to be crucial to interpret the cooling and contraction of fluid planets \citep{Guillot+Showman2002, Rauscher+Showman2014}, their phase curves \citep{Showman+Guillot2002, Knutson+2007, Rauscher+Menou2010}, chemistry and cloud structure \citep{Stevenson+2010, Parmentier+2016, Ehrenreich+2020}. The possibility to study temperate planets will add another layer of complexity due to the additional time variability introduced by storms powered by condensation, as already seen in the case of cool brown dwarfs \citep{Apai+2017}. Information on the spatial distribution of these structures can be retrieved \citep{Crossfield2014, Rauscher+2018} but spatial information will remain very limited. In this context, having the possibility to validate cloud ensemble models and GCMs for hydrogen atmospheres against detailed observations of solar system giant planets, in particular those for Uranus or Neptune, appear essential for further progress.  



\vspace{-12pt}
\section{Keys to the formation of giant planets}
\vspace{-12pt}
Several planetary embryos of sizes comparable to those of Uranus and Neptune may have existed even when Jupiter and Saturn had already reached their final mass \citep[e.g.,][]{Izidoro+2015}. Planets of similar masses and/or radii abound in the Universe \citep{Fulton+2017}, and can start to be characterized \citep[e.g.,][]{Kreidberg+2014}. But rather than a connection based on the mass or sizes of the planets, a key to an Uranus or Neptune mission is that the findings apply to all planets with hydrogen atmospheres, particularly those for which we expect molecular weight gradients to be an important part of their structure and evolution, such as super-Earths with hydrogen rich atmospheres \citep[e.g.][]{Miller-Ricci+2009, Ikoma+Hori2012}. Knowing how heat and chemicals are transported in Uranus and Neptune's atmospheres will provide us with the tools to interpret future spectra of spatially unresolved exoplanets with hydrogen atmospheres.  

\begin{figure*}
\centering{\includegraphics[width=14.0cm]{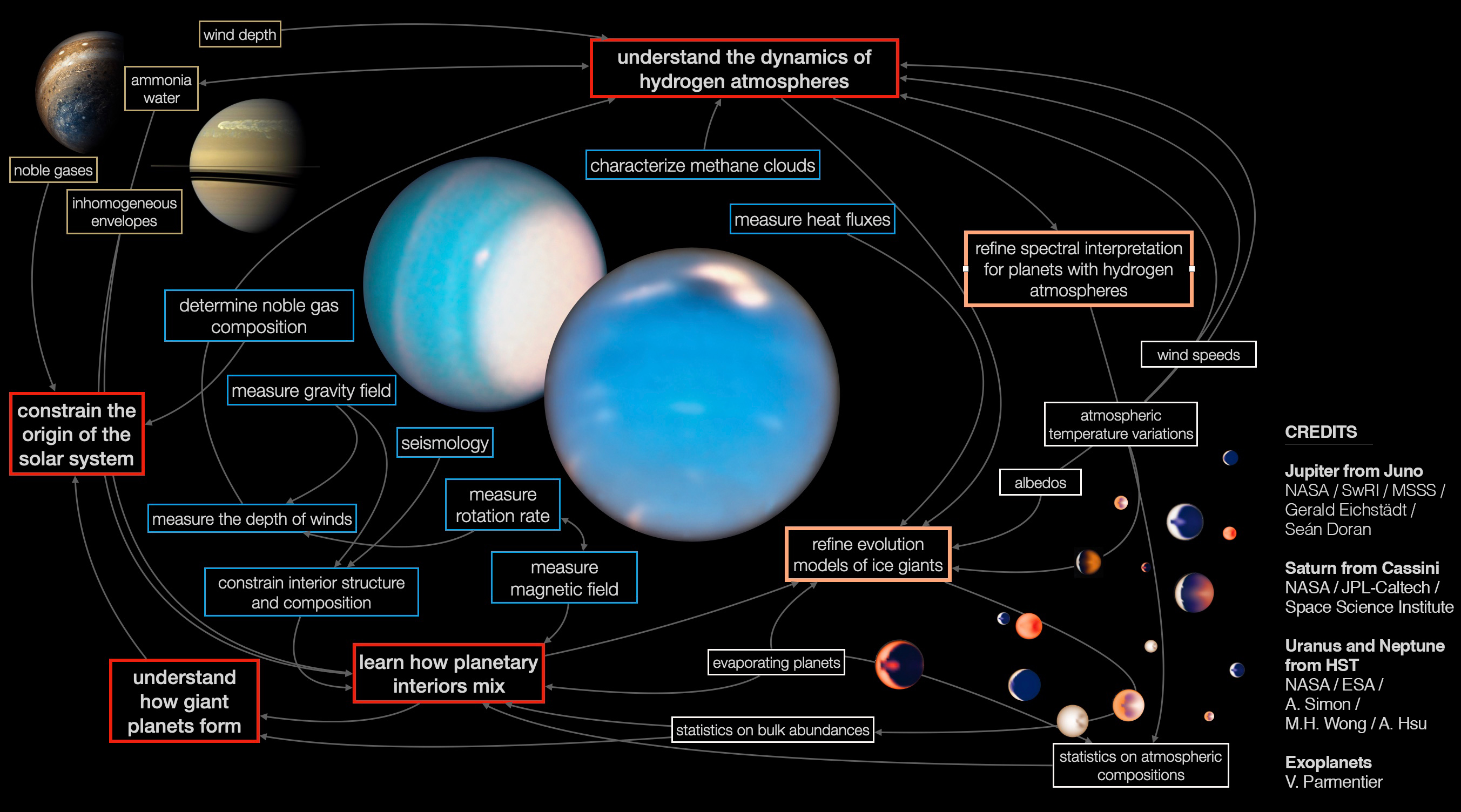}} 
\caption{Understanding the compositions, interior structure, evolution and formation of giant planets requires information from many different sources. The exploration of Uranus and Neptune provides essential pieces of that puzzle, bridging a gap between gas giants Jupiter and Saturn and exoplanets. 
}
	\label{fig:global} 
\end{figure*}

As shown in Fig.~\ref{fig:global}, understanding the formation of giant planets requires combining information obtained from different approaches. Missions around Jupiter and Saturn such as Juno and Cassini have lifted some of the veils on the complexity of the atmospheres of these planets and of their deep structure, including the presence of deep zonal flows \citep[e.g][]{Kaspi+2018, Guillot+2018, Iess+2019, Galanti+2019}, inhomogeneities of their envelopes \citep{Wahl+2017, Debras+Chabrier2019, Ni2019}, and evidence for stable regions \citep{Fuller2014}. But, as discussed in Section~2, heat transport in the presence of condensates remains poorly understood. Observations of exoplanets will provide statistical information on global compositions, wind speeds, variability, but will lack the details that are available for the solar system planets. These details are crucial to constrain simulations of atmospheric dynamics that can then be applied to non-resolved exoplanets. This will become particularly important for temperate exoplanets (in particular when water condenses, because of its role in fueling storms) and for ice giants (due to the higher abundances of heavy elements and higher degeneracy in the interior structure).  

Detailed characterization of Uranus or Neptune by an orbiter and a planetary probe is key. First, the determination of atmospheric dynamics fueled by abundant methane condensation will be crucial to determine the frequency, depth and temperature profiles associated to convective events. Variations of abundances and temperatures in latitude and longitude (for example associated to large-scale circulation, vortex formation and waves) and their variability should be determined. This will be particularly important to determine whether the deep atmosphere is relatively homogeneous in entropy and validate (or not) the 1D approach to the interior structure.

Questions of the deep interior structure, magnetic field, and rotation rate are also paramount to our understanding. The planetary rotation rates are in question \citep{Helled+2010}, interior models are extremely poorly constrained \citep{Nettelmann+2013}, only an upper limit on wind depth was determined \citep{Kaspi+2013} and we do not know where the magnetic field is generated \citep{Stanley+Bloxham2006,Soderlund+2013}. These can all be addressed by precise measurements of the gravity and magnetic fields.  


The evolution of Uranus and Neptune themselves, with Uranus having an order of magnitude smaller intrinsic heat flux than Neptune \citep{Pearl+Conrath1991} remains a mystery. We do not have the solution, but it certainly requires a complete understanding of heat transfer in these planets' atmospheres. Being able to better spot the difference in internal structures of Uranus and Neptune, as determined from the measurement of their gravitational moments and magnetic fields will be crucial. 

Finally, some measurements performed in Uranus and Neptune can help reconstruct the history of the formation of the Solar System. Noble gases, which cannot be seen via remote sensing, are particularly important because they could only be trapped at very low temperatures in the protosolar disk. Their abundance in the atmospheres of Uranus and Neptune compared to that in Jupiter would be an essential piece of the puzzle to determine e.g. whether photoevaporation in the late solar system or clathrate formation may have taken place \citep{Guillot+Hueso2006, Monga+Desch2015, Mousis+2019}. 

The knowledge gained in understanding Uranus and Neptune can be directly applied to known exoplanets. It will also be essential to understand the early stages of planet formation, when planetary embryos should possess a hydrogen atmosphere that is polluted with heavy elements.  In particular water, ammonia and methane are expected to have a large impact on the cooling and the final properties of these forming planets \citep{Kurosaki+Ikoma2017}. Combining knowledge obtained for Jupiter, Saturn, and numerous exoplanets to the information gained from a mission to Uranus or Neptune will allow a complete picture to understand planets with hydrogen atmospheres.

\vspace{-12pt}
\section{Conclusion}
\vspace{-12pt}
Uranus and Neptune hold some of the keys to understand planets with hydrogen atmospheres, finalize the inventory of the Solar System, and infer the history of its formation. A dual mission with an orbiter and a probe, to either planet, reaching all the objectives described in this proposal would be best achieved through an international collaboration.  The experience with Juno has shown that this may be possible within the New Frontier cap and we thus encourage NASA and other partners such as ESA and JAXA to develop a process by which they can partner more easily. Such a mission will be a much awaited milestone in the exploration of our Solar System and will provide the tools needed for the interpretation of observations of planets in our Galaxy.

\vspace{-15pt}


\footnotesize
\printbibliography[title={Bibliography}] 


\end{document}